\documentstyle [12pt] {article}

\begin{document}

\setlength{\baselineskip}{7.5mm}

\begin{flushright}
ITP-SB-92-5
\end{flushright}

\begin{center}
{\Large \bf  New Tools for Analyzing Quark Mixing}  \\

\vspace{20mm}

Alexander Kusenko\footnote{Supported in part by NSF contract
PHY-89-08495 }  \\
Institute for Theoretical Physics  \\
State University of New York   \\
Stony Brook, NY  11794-3840\footnote{ email: kusenko@sbnuc1.phy.sunysb.edu} \\

\vspace{16mm}

{\bf Abstract}

\end{center}

\begin{small}
We present some new mathematical tools which help to derive information about
the quark mass matrices directly from experimental data and to elucidate the
structure of these mass matrices.
\end{small}

\vspace{70mm}

\pagebreak

  One hopes that a successful choice of a mass matrix might help to suggest
an appropriate dynamical model explaining the quark mass spectrum.
In any case, one has to be able to write a mass matrix consistent with the
experiment, regardless of what particular dynamics are responsible for
generating quark masses.  Over the years, a number of proposals have been
advanced for quark mass matrices, such as those in refs.
\cite{Fritzsch,Kauss,Albright,Branco1,Gupta,Shrock}.
In this paper we would like to present
some useful mathematical tools for the study of quark mass matrices and the
connection with quark mixing.  The starting point of our study is an
observation which has so far not been exploited: since the
Cabibbo-Kobayashi-Maskawa (CKM) quark mixing matrix $V$ is an element of the
Lie group U(3), it is in one-to-one correspondence with an element $ H $
of the Lie algebra $ u(3) $.  $H$ is thus the infinitesimal generator for $V$.
It is of interest to determine the $H$ corresponding to the empirically
measured $V$.  As we shall show, one can also use $H$ to facilitate the
determination of families of quark mass matrices which are consistent with
experiment.  (Here we use the fact \cite{data} that there are three light
neutrinos, hence three usual SU(2) doublets of leptons, and, as required,
e.g., by anomaly cancellation, thus also three SU(2) doublets of quarks, so
that $V \in U(3)$.)

We denote the quark mass matrices for the up and down sectors $ M_{u}$
and $ M_{d} $ correspondingly.
By means of a bilinear transformation one can diagonalize $ M_{u}$ and
$ M_{d} $:

\begin{eqnarray}
\left \{ \begin{array}{l}
        U_{u,L} M_{u} U_{u,R}^{\dag}=diag(m_{u},m_{c},m_{t}) \equiv D_{u}  \\
        U_{d,L} M_{d} U_{d,R}^{\dag}=diag(m_{d},m_{s},m_{b}) \equiv D_{d}  \\
         \end{array}    \right.
\end{eqnarray}
where $ U_{u,L}, U_{d,L}, U_{u,R}, U_{d,R} $ are unitary matrices.
Below we shall suppress the subscript $ L $.

  In the standard model with only left-handed charged
weak currents, the matrix $ U_{R} $ is not a measurable quantity; and therefore
one may choose the $ U_{u,R}$ and $ U_{d,R} $
in such a way that $ M_{u} $ and $ M_{d} $ are
both hermitian.  Indeed, an arbitrary complex $ 3 \times 3 $ matrix
is defined by 18 real parameters.  In order to make it hermitian,
one has to impose 9 constraints to eliminate 9 degrees of freedom. To do that
one may use 9 degrees of freedom associated with each of the matrices
$ U_{u,R} $ and  $ U_{d,R} $.  Below, we  assume both $ M_{u}$ and $ M_{d} $
are hermitian.

  The experimental data \cite{data}
provides one with the absolute values of the matrix elements of the CKM
mixing matrix
$ V =U_{u} U_{d}^{\dag} $ (taken to be $3 \times 3$).
One also obtains certain phase information from experiment which, for a
given parametrization, can
be translated into a determination of the phases of the elements of $V$.
  Suppose the pair of hermitian mass matrices $ M_{u} $ and $ M_{d} $,

\begin{eqnarray}
\left \{ \begin{array}{l}
M_{u}=U_{u}^{\dag} D_{u} U_{u}   \\
\\
M_{d}=U_{d}^{\dag} D_{d} U_{d}
         \end{array}
\right .
\end{eqnarray}
corresponds to some particular matrix $ V $.  Then one can see that for every
unitary matrix $ U_{0} $ there exists another  pair of hermitian matrices:

\begin{eqnarray}
\left \{ \begin{array}{l}
M_{u}'=(U_{u} U_{0})^{\dag} \ D_{u} \ (U_{u} U_{0})   \\
\\
M_{d}'=(U_{d} U_{0})^{\dag} \ D_{d} \ (U_{d} U_{0})
         \end{array}
\right .
\end{eqnarray}
that corresponds to the same matrix $ V $.  One can also see that the family
of matrices $ M_{u}'$ and $ M_{d}'$ as a function of an
arbitrary unitary matrix
$ U_{0} $ exhausts all possible solutions for the mass matrices.  This proves
that for every mixing matrix $ V \in U(3) $ there
exists a 9-parameter family of matrices that might be chosen to be mass
matrices.

Empirically, the data constrains $ | V | $, the matrix of magnitudes
of each of the elements of $ V $, to be close to the
identity matrix.
Therefore, it is useful to represent $ V $ as

\begin{equation}
V=e^{i \alpha H}=U_{u} U_{d}^{\dag}   \label{exp}
\end{equation}
where $ H $ is some hermitian matrix and $ \alpha $ is a real number.
One may choose $ H $ to have its dominant (largest, in absolute value)
eigenvalue to be of the order of $ 1 $, yet smaller than $ 1 $.
Then for $ \alpha $ consistent with the data \cite{data}, we find that
$ | \alpha | \approx 0.3 $.  We will use the fact that
$ \alpha < 1$ below.

  Suppose that $ V $ is known. Six parameters, the phases
of the quark fields,
are arbitrary.  We use three of them to rephase $ V $ to make it
be close to the $ 3 \times 3 $ identity matrix.  This is possible since
$ | V | \approx {\bf 1} $.
Then one can calculate the matrix $ \alpha H $ using the Sylvester's
theorem \cite{matrix}:

\begin{equation}
i \alpha H=\sum_{k=1}^{3} ln(v_{k})
\frac{\prod_{i \neq k}(V-v_{i} \times {\bf 1})}
{\prod_{i \neq k} (v_{k}-v_{i})}    \label{Sylvester}
\end{equation}
The latter is valid as long as all the eigenvalues of $ V $ are distinct.
  If the matrix $ V $ has 2 or 3 degenerate eigenvalues, then  $ ln(V) $ is
not uniquely defined, and there is a
family of matrices $ H $ that satisfy (\ref{exp}).  This family of solutions
is explicitly shown in \cite{matrix}.  However, this mathematical subtlety is
not of physical importance since
one can always avoid the degenerate case by some small change in the
phases of quark fields that removes any initial degeneracy in eigenvalues.
The theorem (\ref{Sylvester}) is particularly powerful since it
gives an exact result in a closed form.

We will demonstrate how one could construct a set of mass matrices starting
from the mixing matrix $ V $.  In particular, we choose to describe those
mass matrices that can be diagonalized by a unitary transformation
$ U_{u;d} $ generated
by an element in the $ u(3) $ algebra that commutes with $ H $.

In terms of the matrix $ H $,

\begin{eqnarray}
V={\bf 1}+ i \alpha H- \frac{1}{2} \alpha^{2} H^{2} +...
+\frac{1}{n!}(i\alpha H)^{n}+...
\end{eqnarray}

One can also decompose an arbitrary $ U_{1} $ and $ U_{2} $, that commute
with $ H $, in a similar series whose coefficients are proportional to certain
real numbers: $ \{ x_{n}^{(u)} \} $ and
$ \{ x_{n}^{(d)} \} $ where $ n=1,2,3...$

\begin{eqnarray}
\left \{ \begin{array}{l}
U_{u}={\bf 1}+i \alpha x_{1}^{(u)} H- \frac{1}{2} x_{2}^{(u)} \alpha^{2} H^{2}
+...+\frac{1}{n!} x^{(u)}_{n} (i \alpha H)^{n}+... \\
\\
U_{d}={\bf 1}+i \alpha x_{1}^{(d)} H- \frac{1}{2} x_{2}^{(d)} \alpha^{2} H^{2}
+...+\frac{1}{n!} x^{(d)}_{n} (i \alpha H)^{n}+...
         \end{array}
\right.
\end{eqnarray}
\noindent
This series is convergent if

\[ \sup_{n=1,2,3...} | x_{n} | < \infty     \]

\noindent
The constraints $ U_{u} U_{d}^{\dag}=V, \ U_{u}^{\dag} U_{u}={\bf 1} $ and
$ U_{d}^{\dag} U_{d}={\bf 1} $ are satisfied simultaneously up to the third
order in $ \alpha $ if the coefficients are chosen as follows:

\begin{eqnarray}
\left \{ \begin{array}{l}
 U_{u}={\bf 1} + i \alpha H x- \frac{\alpha^{2}}{2} H^{2} x^{2} +...  \\
\\
 U_{d}={\bf 1} + i \alpha H (x-1)-
\frac{\alpha^{2}}{2} H^{2} (x-1)^{2} +...
         \end{array}  \right.
\end{eqnarray}
where $ x $ is some real parameter.
Then, to the second order in $ \alpha $, one may express $ M_{u} $ and
$ M_{d} $ as functions of $ x $:

\begin{eqnarray}
\left \{ \begin{array}{l}
M_{u} =U_{u}^{\dag} D_{u} U_{u} =      \\  \\
D_{u}+i \alpha x [D_{u},H]-\frac{1}{2} \alpha^{2} x^{2} [[D_{u},H],H]+...
\\  \\                                              \\  \\
M_{d} =U_{d}^{\dag} D_{d} U_{d}=       \\  \\
D_{d}+i \alpha (x-1) [D_{d};H]-
\frac{1}{2} \alpha^{2} (x-1)^{2} [[D_{d},H],H]+...
         \end{array}    \right.
\end{eqnarray}

We see that, in our parametrization, one gets a good approximation of $  M $'s
as a function of one real parameter.  Thus, from $V$ and the resultant $H$, we
have obtained a family of (experimentally acceptable) mass matrices depending
on a parameter $x$.

In passing, we note the following technical point: for rapid convergence of the
series (9), $ V $ should be close to $ {\bf 1} $, the identity.
The three phases of the diagonal elements of $ V $
are not physical and they may be changed by rephasing the quark fields.
Such a change in the overall phase of $ V $ corresponds to shifting
$ \alpha $ by a real number.  Clearly,
$ | \alpha | $ is the smallest for $ V \approx {\bf 1} $, which yields rapid
convergence of both (6) and (9).

The family of solutions (9) for
the ``up'' and ``down'' quark
mass matrices (accurate to second order in $\alpha$) is

\begin{eqnarray}
\left \{ \begin{array}{l}
M_{u} \approx D_{u}+i \alpha x [D_{u},H]-\frac{1}{2} \alpha^{2} x^{2}
[[D_{u},H],H]
             \\                                 \\
M_{d}  \approx D_{d}+i \alpha (x-1) [D_{d},H]
-\frac{1}{2} \alpha^{2} (x-1)^{2} [[D_{d},H],H]
         \end{array}    \right.
\end{eqnarray}

  To illustrate the method with a simple numerical example let us take

\begin{eqnarray}
V=\left ( \begin{array}{ccc}
0.97525 &  0.22104 & 0.0050614 e^{-i \ 40.00^{o}}        \\
-0.22099 e^{i 0.0359^{o}} & 0.97430 e^{-i \ 0.00184^{o}} &  0.043619      \\
0.0066657 e^{-i \ 28.40 ^{o}} & -0.043403 e^{i \ 0.9485^{o}}  & 0.99904
          \end{array}
\right )
\end{eqnarray}
(consistent with current data \cite{data}).  Note that this gives
$ |V_{ub}|/|V_{cb}| = 0.11 $ and, for the parametrization-invariant CP
violation quantity \cite{Jarlskog} $J=
Im \{V_{11}V_{22}V_{12}^{\dagger}V_{21}^{\dagger}\}$ the value
$J= 3.1 \times 10^{-5}$.

  First, we calculate the eigenvalues of $ V $:
$ v_{1}=0.97415+0.22587 i, \ v_{2}=0.97443-0.22468 i,
\ v_{3}=0.999999+0.00122 i $.  Next, we substitute these into
(\ref{Sylvester}) to calculate $ H $ and obtain:

\begin{eqnarray}
\alpha H=\left ( \begin{array}{ccc}
 0        & -0.223 i        & -0.0032-0.001 i  \\
 0.223 i  &  0              & -0.044 i   \\
 -0.0032+0.001i  & 0.044 i    & 0
                  \end{array}
\right )
\end{eqnarray}

We now choose the quark masses to be equal to their running masses evaluated
at $ Q_{0}=1 \ GeV $ and with $ \Lambda =200 \ MeV $ in the
$ \bar{MS} $ renormalization scheme.  From the analysis of \cite{masses}:
$ \bar{m}_{u}(Q_{0})=(5.1 \pm 1.5) \ MeV,
\ \bar{m}_{d}(Q_{0})=(8.9 \pm 2.6) \ MeV , \
\bar{m}_{s}(Q_{0})=(175 \pm 55) \  MeV, \
\bar{m}_{c}(Q_{0})=(1.35 \pm 0.05) \ GeV,\
\bar{m}_{b}(Q_{0})=(5.9 \pm 0.1) \ GeV. $

We also use the value $ \bar{m}_{t}(Q_{0})=220 \ GeV $ corresponding to a
physical mass of $ m_{t}^{phys}=130 \ GeV $, consistent with current
constraints.  Specifically, we choose the diagonal mass matrices to be

\begin{eqnarray}
\begin{array}{l}
D_{u}=diag(5 \ MeV, \ -1.35 \ GeV, \ 220 \ GeV)     \\
D_{d}=diag(8.9 \ MeV, \ -175 \ MeV , \ 5.9 \ GeV)
\end{array}
\end{eqnarray}

  Suppose $ x=1/2 $.  In this case, one retains a symmetry between the
``up'' and ``down'' sectors in (15).  For these values of quark masses the
first order approximation of the ``up'' and ''down'' sector mass matrices
gives the following:

\begin{eqnarray}
M_{u}= 220 GeV  \left ( \begin{array}{ccc}
-5.09\times 10^{-5} & (6.77 \times 10^{-4}) e^{-i \ 3^{o}} &
(2.37 \times 10^{-3}) e^{i \ 42^{o}}
  \\
(6.77 \times 10^{-4}) e^{i \ 3^{o}} &
-5.58 \times 10^{-3}     & -0.0221   \\
(2.37 \times 10^{-3}) e^{-i \ 42^{o}}  & -0.0221   &  0.99951
              \end{array}
\right )
\end{eqnarray}

\begin{eqnarray}
M_{d}=5.9 GeV  \left ( \begin{array}{ccc}
1.49 \times 10^{-3} & -3.47 \times 10^{-3}
& (1.78 \times 10^{-3})e^{-i \ 66^{o}} \\
-3.47 \times 10^{-3}   &   -0.0293    & 0.0226    \\
 (1.78 \times 10^{-3})e^{i \ 66^{o}} & 0.0226     &  0.99951
              \end{array}
\right )
\end{eqnarray}

Even though this is only a second order
approximation it still shows what
is the relative value of different terms in the mass matrices for this
particular case of $ x=(1/2) $.  For instance, the matrix $ V $
calculated back from $ M_{u} $ and $ M_{d} $
is different from the original $ V $ only in the third significant digit.

It is of interest to express the CP violation quantity $J$ directly
in terms of $H$.  Recall that

\begin{equation}
J=-\frac{1}{2 F_{u} F_{d}} det(C)  \label{def}
\end{equation}
where
\begin{eqnarray}
                \begin{array}{l}
F_{u}=(m_{t}-m_{c})(m_{t}-m_{u})(m_{c}-m_{u}) \\
F_{d}=(m_{b}-m_{s})(m_{b}-m_{d})(m_{s}-m_{d})
                \end{array}
\end{eqnarray}

\begin{equation}
C = -i [ M_{u},M_{d}]     \label{C-def}
\end{equation}

\noindent Now

\begin{eqnarray}
   \left\{     \begin{array}{l}
 M_{u}= U_{u}^{\dag}D_{u}U_{u}=U_{d}^{\dag}V^{\dag}  D_{u} V U_{d}=  \\
\\
U_{d}^{\dag} ( D_{u}+i\alpha [D_{u},H]-\frac{\alpha^{2}}{2}[[D_{u},H],H]+...)
U_{d}  \\
     \  \\
M_{d}=U_{d}^{\dag} D_{d} U_{d}
                \end{array}
\right.
\end{eqnarray}

\noindent
so that up to the first non-vanishing order in $ \alpha $,

\begin{equation}
iC \equiv [ M_{u}, M_{d}]=
i (\alpha+O(\alpha^{2})) \
U_{d}^{\dag} [[H,D_{u}],D_{d}] U_{d}  \label{C-approx}
\end{equation}

Using (\ref{C-approx}), we thus find

\begin{equation}
J=(\alpha+O(\alpha^{2}))^{3}
Re[H_{12} H_{23} H_{31}]          \label{J}
\end{equation}
\noindent

This result allows one to calculate the parameter $ J $ up to the small
corrections of the order of $ \alpha^{2} $.  The higher order terms
can be obtained by retaining the corresponding higher order terms
in (\ref{C-approx}).
For example, for the numerical illustration (11)-(15),
(\ref{J}) yields $J \simeq 3.2
\times 10^{-5}$, very close to the exact value $J=3.1 \times 10^{-5}$.

   These results thus establish a useful connection between the $ M $
matrices, the unitary transformations which diagonalize them, and the actual
quark mixing matrix $ V $.  Furthermore, they introduce an
 interesting object, the matrix $ H $ in Lie algebra u(3)
corresponding to $ V $ in the Lie group U(3).  We have shown how with our
tools, one can obtain a family of experimentally acceptable mass matrices
directly, given knowledge of $V$.  This technique complements the earlier
typical one of hypothesizing some ansatz for a mass matrix and then checking to
see if it fits experiment.

\vspace{7mm}

The author would like to thank Professor Robert E. Shrock for many helpful
discussions and comments.

\end{document}